\documentclass[aps,twocolumn,prd,showpacs,showkeys,preprintnumbers,superscriptaddress,nobibnotes,floatfix,longbibliography,nofootinbib]{revtex4-2}
\pdfoutput=1
\usepackage{amsmath}
\usepackage{amsfonts}
\usepackage{amssymb}
\usepackage{mathrsfs}
\usepackage{graphicx}
\usepackage{color}
\usepackage{hyperref}
\usepackage{multirow}
\usepackage{slashed}
\usepackage{epsf}

\usepackage{graphicx}
\usepackage{caption}
\usepackage{subcaption}

\usepackage{color}

\def\to{\rightarrow}

\begin{document}

\title{
The Light Neutralino Dark Matter at Future Colliders in the MSSM with the Generalized Minimal Supergravity (GmSUGRA) }

\author{Imtiaz Khan}
\email{ikhanphys1993@gmail.com}
\affiliation{Department of Physics, Zhejiang Normal University, Jinhua, Zhejiang 321004, China}
\affiliation{Zhejiang Institute of Photoelectronics, Jinhua, Zhejiang 321004, China}

\author{Ali Muhammad}
\email{alimuhammad@phys.qau.edu.pk}
\affiliation{CAS Key Laboratory of Theoretical Physics, Institute of Theoretical Physics, Chinese Academy of Sciences, Beijing 100190, China}
\affiliation{School of Physical Sciences, University of Chinese Academy of Sciences, No. 19A Yuquan Road, Beijing 100049, China}

\author{Tianjun Li}
\email{tli@itp.ac.cn}
\affiliation{CAS Key Laboratory of Theoretical Physics, Institute of Theoretical Physics, Chinese Academy of Sciences, Beijing 100190, China}
\affiliation{School of Physical Sciences, University of Chinese Academy of Sciences, No. 19A Yuquan Road, Beijing 100049, China}
\affiliation{School of Physics, Henan Normal University, Xinxiang 453007, P. R. China}

\author{Shabbar Raza}
\email{shabbar.raza@fuuast.edu.pk}
\affiliation{Department of Physics, Federal Urdu University of Arts, Science and Technology, Karachi 75300, Pakistan}

\author{Pirzada}
\email{pirzadawaqar60@gmail.com}
\affiliation{CAS Key Laboratory of Theoretical Physics, Institute of Theoretical Physics, Chinese Academy of Sciences, Beijing 100190, China}
\affiliation{School of Physical Sciences, University of Chinese Academy of Sciences, No. 19A Yuquan Road, Beijing 100049, China}

\author{Mussawir Khan}
\email{mussawirkhan@ihep.ac.cn}
\affiliation{State Key Laboratory of Particle Astrophysics, Institute of High Energy Physics, Chinese Academy of Sciences, Beijing 100049, China}
\affiliation{University of Chinese Academy of Sciences, Beijing 100049, China}

\begin{abstract}
We perform a detailed investigation of light right-handed slepton bulk regions, together with the Higgs- and $Z$-resonance regimes, in the Minimal Supersymmetric Standard Model (MSSM) with the Generalized Minimal Supergravity (GmSUGRA), focusing on the higgsino mass parameter scenario $\mu < 0$, given that the anomalous magnetic moment of the muon may now be consistent with the Standard Model (SM) prediction. A systematic numerical exploration of the parameter space is carried out, where the bulk region is conservatively defined by the mass-splitting ratio $\mathcal{R}_{\tilde{\phi}} = (m_{\tilde{\phi}} - m_{\tilde{\chi}_1^0})/m_{\tilde{\chi}_1^0} \gtrsim 10\%$. In particular, we analyze the case in which the right-handed stau ($\tilde{\tau}_R$) emerges as the Next-to-Lightest Supersymmetric Particle (NLSP). We uncover a sizeable parameter space consistent with current experimental bounds, including limits from the LHC supersymmetry searches, the Planck 2018 relic density measurement, and direct detection constraints on neutralino–nucleon scattering. This region naturally accommodates bulk annihilation channels mediated by right-handed sleptons. We obtain robust upper bounds on the masses of the lightest neutralino, $m_{\tilde{\chi}_1^0} \lesssim 143~\text{GeV}$, and the right-handed stau, $m_{\tilde{\tau}_R} \lesssim 158~\text{GeV}$. The identified bulk region lies within the prospective reach of forthcoming dark matter direct detection facilities such as LUX-ZEPLIN, as well as future high-energy $e^+e^-$ colliders including FCC$_{\rm ee}$ and CEPC. Conversely, the scenarios with the right-handed selectron as the NLSP have already been excluded by the current LHC data. Furthermore, the parameter space consistent with our findings yields contributions to the anomalous magnetic moment of the muon, $a_\mu = (g_\mu -2)/2$, that remain within $1\sigma$ of the experimental central value, thereby reinforcing the phenomenological viability of supersymmetric dark matter in this framework.
\end{abstract}

\maketitle


Supersymmetric extensions of the Standard Model (SM) continue to be among the most promising theoretical frameworks for addressing phenomena Beyond the Standard Model (BSM). Despite the absence of direct experimental confirmation, the supersymmetric standard models (SSMs) continue to attract interest due to their remarkable theoretical virtues. Chief among these are the natural resolution to the gauge hierarchy problem, the gauge coupling unification of  at high scales~\cite{gaugeunification,Georgi:1974sy,Pati:1974yy,Mohapatra:1974hk,Fritzsch:1974nn,Georgi:1974my}, and the prediction of the lightest supersymmetric particle (LSP) is stable and weakly interacting, making it a well-motivated candidate for cold dark matter (DM) candidate in $R$-parity conserving scenarios~\cite{neutralinodarkmatter,darkmatterreviews}. Furthermore, supersymmetry (SUSY) facilitates radiative electroweak (EW) symmetry breaking, primarily driven by the large top-quark Yukawa coupling, thereby establishing a predictive connection between high-scale physics and observable phenomena at low energies. 

One of the significant achievements of the Minimal Supersymmetric Standard Model (MSSM) is its prediction that the mass of the lightest CP-even Higgs boson should lie within the interval $100$--$135$ GeV~\cite{Slavich:2020zjv}.This range aligns well with the mass of the Higgs boson observed at the Large Hadron Collider (LHC), lending support to the model’s phenomenological viability. This has reinforced the MSSM as a prominent framework for new physics. Nonetheless, after Run-2 of the LHC, no ultimate signal of supersymmetric particles has emerged. The latest collider data have established robust exclusion bounds, pushing the minimum allowed masses for stop and sbottom quarks to about $1.25$ TeV and $1.5$ TeV, respectively, while first- and second-generation squarks are excluded below $2$ TeV and gluinos below $2.2$  TeV~\cite{ATLAS-SUSY-Search, Aad:2020sgw, Aad:2019pfy, CMS-SUSY-Search-I, CMS-SUSY-Search-II}. In consequence, colored sparticles must lie in the multi-TeV ranges, rendering them inaccessible to present collider energies. 

In contrast, an electroweak-scale bino-like neutralino LSP remains phenomenologically viable. However, the heavy sfermion spectrum tends to suppress the neutralino pair-annihilation cross section mediated by sfermion exchange. In the absence of additional mechanisms, this suppressed interaction leads to a relic abundance that is higher than the dark matter density observed by the Planck satellite, $0.114 \leq \Omega_{\rm CDM} h^2 \leq 0.126$ (at $5\sigma$)~\cite{Akrami:2018vks,Drees:1992am}. Nonetheless, several model realizations still accommodate the light sleptons with masses of a few hundred GeV to $\mathcal{O}(\mathrm{TeV})$, even when squarks are substantially heavier~\cite{Ahmed:2022ude, Zhang:2023jcf, Khan:2023ryc}. 

The insufficiency of bino-dominated annihilation has motivated several well-studied mechanisms to bring the DM relic density into agreement with observation: 
\begin{itemize}
    \item \textbf{Bulk region:} If one of the sfermion is sufficiently light, neutralino annihilation into SM fermions via $t$- and $u$-channel sfermion exchange is efficient enough to deplete the relic density without invoking additional mechanisms~\cite{King:2006tf}.
    \item \textbf{Higgs/Z funnels:} When $m_{\tilde{\chi}_1^0} \simeq m_h/2$ or $m_Z/2$, neutralinos annihilate resonantly through the Higgs or $Z$ pole, dramatically enhancing the annihilation cross section.
    \item \textbf{Coannihilation:} If the mass of a sfermion is nearly degenerate with the neutralino LSP, coannihilation channels further reduce the relic abundance.
    \item \textbf{Well-tempered scenario:} A neutralino with a significant higgsino or wino admixture annihilates efficiently through gauge interactions, naturally satisfying the relic density bounds. 
\end{itemize}

The bulk region scenario provides a well motivated framework for dark matter, as it naturally achieves the correct relic abundance through standard annihilation processes without relying on precise mass alignments or resonant enhancements.  A particularly intriguing case arises when $m_{\tilde{\chi}_1^0} \lesssim m_h/2$, enabling the invisible Higgs decay $h \rightarrow \tilde{\chi}_1^0 \tilde{\chi}_1^0$, which links the Higgs sector directly to the DM candidate. This possibility has been actively pursued in experimental searches for invisible Higgs decays~\cite{ATLAS:2022yvh}, complementing collider probes of heavy Higgs bosons~\cite{ATLAS:2020zms}, electroweakinos~\cite{CMS:2020bfa, ATLAS:2021moa, ATLAS:2021yqv, CMS:2022sfi}, and recent direct detection (DD) constraints from XENONnT and LUX-ZEPLIN (LZ)~\cite{XENON:2023cxc,LZ:2018qzl,LZ:2022lsv,LZ:2024zvo}. The latest LZ bounds in particular have set unprecedented limits on spin-independent neutralino-nucleon scattering, motivating a reappraisal of light neutralino parameter space. 

To reconcile these experimental bounds while retaining natural supersymmetric realizations, the concept of \emph{Electroweak Supersymmetry} (EWSUSY) was proposed~\cite{Cheng:2012np, Cheng:2013hna, Li:2014dna}. In this framework, squarks and gluinos are heavy, but electroweak sparticles (bino, wino, sleptons, and sneutrinos) remain near the electroweak scale. This mass pattern can be realized naturally within the Generalized Minimal Supergravity (GmSUGRA) framework~\cite{Li:2010xr, Balazs:2010ha}, simultaneously satisfying collider limits and DM relic density constraints. 

The present work focuses on the bulk region and Higgs/$Z$ funnels in the MSSM within the GmSUGRA framework, with particular emphasis on the case $\mu < 0$, given that recent theoretical calculations suggest the SM prediction for the muon anomalous magnetic moment could align with experimental measurements. While a positive value of $\mu$ has been traditionally preferred to explain the experimental anomaly in the muon anomalous magnetic moment $a_\mu = (g-2)_\mu/2$~\cite{Muong-2:2023cdq, Ahmed:2021htr, Muong-2:2021ojo}, uncertainties in the hadronic contributions render the sign of $\mu$ an open question. Interestingly, scenarios with $\mu < 0$ have been shown to make distinctive contributions to $a_\mu$~\cite{Khan:2025yit}, and the sign also plays a decisive role in shaping Higgs/$Z$ resonance solutions~\cite{Khan:2025azf}. 

In exploring the bulk region within GmSUGRA, we restrict attention to right-handed sleptons, as other sfermions are necessarily heavy given the current LHC bounds. The nature of DM annihilation versus coannihilation is governed by the mass splitting between the LSP and the light right-handed sleptons, conveniently quantified by $\mathcal{R}_{\tilde{\phi}} \equiv ({m_{\tilde{\phi}} - m_{\tilde{\chi}_1^0}})/{m_{\tilde{\chi}_1^0}} \,,$
where $\tilde{\phi}$ denotes $\tilde{e}_R$ or $\tilde{\tau}_1$. A conservative requirement for pure annihilation-dominated dynamics (excluding coannihilation and resonance effects) is $\mathcal{R}_{\tilde{\phi}} \gtrsim 10\%$. Our analysis demonstrates that in scenarios where the $\tilde{\tau}_1$  state is the secondary lightest superpartner, a significant fraction of the model's allowed configurations remains viable. This region respects the Planck relic density constraints, LHC limits, and the latest Direct Detection (DD) bounds, with the upper mass bounds of $m_{\tilde{\chi}_1^0} \simeq 145$ GeV and $m_{\tilde{\tau}_R} \simeq 160$ GeV. While inaccessible at present LHC energies, these configurations can be probed at the next-generation DD experiments such as LZ, and at future $e^+e^-$ colliders such as the FCC$_{\rm ee}$~\cite{FCC:2018byv, FCC:2018evy} and CEPC~\cite{CEPCStudyGroup:2018ghi}. By contrast, scenarios with a right-handed selectron NLSP has already been excluded by current SUSY searches. 

We also investigate the regime where $m_{\tilde{\chi}_1^0} \lesssim m_h/2$ or $m_Z/2$, enabling invisible decays of the Higgs and $Z$ bosons into neutralinos. The $\mu < 0$ case emerges as especially promising, motivating a systematic re-examination of light neutralino DM in light of present and forthcoming data~\cite{Khan:2025azf}.


\section*{GmSUGRA Framework}
\label{model}

The Generalized Minimal Supergravity (GmSUGRA) framework~\cite{Li:2010xr, Balazs:2010ha} provides a natural realization of EWSUSY~\cite{Cheng:2012np, Cheng:2013hna, Li:2014dna}. In this construction, electroweak sparticles—namely sleptons, neutralinos, and charginos—typically reside at the sub-TeV to TeV scale, while colored states such as squarks and gluinos are naturally heavier, often in the multi-TeV range~\cite{Li:2014dna, Cheng:2012np}. This split spectrum reconciles collider constraints with viable dark matter phenomenology.  

At the grand unification (GUT) scale, GmSUGRA imposes non-trivial relations among gauge couplings and gaugino masses~\cite{Li:2010xr, Balazs:2010ha}:
\begin{equation}
 \frac{1}{\alpha_2}-\frac{1}{\alpha_3} =
 k\left(\frac{1}{\alpha_1}-\frac{1}{\alpha_3}\right),
\end{equation}
\begin{equation}
 \frac{M_2}{\alpha_2}-\frac{M_3}{\alpha_3} =
 k\left(\frac{M_1}{\alpha_1}-\frac{M_3}{\alpha_3}\right).
\end{equation}
In the simplified setup adopted here, the index is fixed to $k=5/3$. Assuming unification of the gauge couplings ($\alpha_1=\alpha_2=\alpha_3$) at the GUT scale, the gaugino mass relation reduces to
\begin{equation}
 M_2 - M_3 = \frac{5}{3}\,(M_1 - M_3).
\label{M3a}
\end{equation}
Thus, the familiar mSUGRA relation $M_1=M_2=M_3$ emerges as a limiting case of GmSUGRA. Rather than three independent gaugino parameters, only two are free, with $M_2$ determined by $M_1$ and $M_3$:
\begin{equation}
 M_2 = \frac{5}{3} M_1 - \frac{2}{3} M_3.
\label{M3}
\end{equation}

The scalar sector in GmSUGRA exhibits a richer structure. While slepton masses are treated as independent inputs, squark masses are generated via $SU(5)$ grand unification with an adjoint Higgs field~\cite{Balazs:2010ha}, leading to the following relations:
\begin{align}
m_{\tilde{q}_i}^2 &= \tfrac{5}{6}(m_0^{u})^2 + \tfrac{1}{6} m_{\tilde{e}_i^c}^2, \\
m_{\tilde{u}_i^c}^2 &= \tfrac{5}{3}(m_0^{u})^2 - \tfrac{2}{3} m_{\tilde{e}_i^c}^2, \\
m_{\tilde{d}_i^c}^2 &= \tfrac{5}{3}(m_0^{u})^2 - \tfrac{2}{3} m_{\tilde{l}_i}^2.
\label{squarks_masses}
\end{align}
Here $m_{\tilde{q}}$, $m_{\tilde{u}^c}$, and $m_{\tilde{d}^c}$ refer to the soft supersymmetry-breaking mass parameters for the left-handed squark doublets, right-handed up-type squarks, and right-handed down-type squarks, respectively. Similarly, $m_{\tilde{l}}$ and $m_{\tilde{e}^c}$ denote the corresponding soft masses for the left-handed and right-handed sleptons.  The parameter $m_0^u$ plays the role of a universal scalar mass, analogous to $m_0$ in mSUGRA.  

In light-slepton realizations, $m_{\tilde{l}}$ and $m_{\tilde{e}^c}$ are restricted to be $\lesssim 1$ TeV. If $m_0^u \gg m_{\tilde{l}}, m_{\tilde{e}^c}$, the squark spectrum approximately satisfies $2 m_{\tilde{q}}^2 \simeq m_{\tilde{u}^c}^2 \simeq m_{\tilde{d}^c}^2$. Finally, within GmSUGRA, the trilinear soft terms $A_u, A_d, A_e$ and the Higgs soft masses $m_{\tilde{H}_u}$ and $m_{\tilde{H}_d}$ are considered as free inputs, offering significant flexibility in model building~\cite{Cheng:2012np, Balazs:2010ha}.

\section*{Methodology: Scanning Procedure, GUT-Scale Parameters, and Phenomenological Constraints}
\label{scan_method}

To investigate the parameters space of the GmSUGRA framework, we employ the ISAJET~7.85 package~\cite{ISAJET}, which incorporates the renormalization group equations (RGEs) of MSSM parameters in the $\overline{DR}$
 regularization scheme. In this approach, the Yukawa couplings of the third-generation fermions are evolved from the electroweak scale up to the grand unification scale $M_{\rm GUT}$. Unlike strict mSUGRA scenarios, we do not impose exact gauge coupling unification ($g_1=g_2=g_3$) at $M_{\rm GUT}$; small deviations, at the level of a few percent, are tolerated as they may arise from unknown GUT-scale threshold corrections~\cite{Hisano:1992jj}. All gauge and Yukawa couplings, along with soft supersymmetry-breaking (SSB) parameters, are evolved from the GUT-scale down to the electroweak scale $M_Z$, by solving the renormalization group equations with the given boundary conditions at the high scale. In ISAJET, the unification point $M_{\rm U}$ is defined by $g_1=g_2$, and two-loop MSSM RGEs are consistently implemented.

The parameter space is probed via a random scan in the following ranges:
\begin{align}
80 \, {\rm GeV} \leq & m_0^{u} \leq 10\,{\rm TeV}, \nonumber \\
0 \, {\rm GeV} \leq & M_1 \leq 1.2\,{\rm TeV}, \nonumber \\
1.0 \, {\rm TeV} \leq & M_3 \leq 3.0\,{\rm TeV}, \nonumber \\
100 \, {\rm GeV} \leq & m_{\tilde l} \leq 5.0\,{\rm TeV}, \nonumber \\
0 \, {\rm GeV} \leq & m_{\tilde e^c} \leq 300\,{\rm GeV}, \nonumber \\
0 \, {\rm GeV} \leq & m_{\tilde H_{u,d}} \leq 10\,{\rm TeV}, \nonumber \\
-10\,{\rm TeV} \leq & A_u=A_d \leq 10\,{\rm TeV}, \nonumber \\
-5.0\,{\rm TeV} \leq & A_e \leq 5.0\,{\rm TeV}, \nonumber \\
2 \leq & \tan\beta \leq 60. 
\label{input_param_range}
\end{align}
We fix the top quark pole mass to $m_t=173.3\,{\rm GeV}$~\cite{:2009ec} and focus on the $\mu<0$ scenario of the Higgsino mass parameter. The results remain stable under small variations of $m_t$ within its $1$–$2\sigma$ uncertainty~\cite{bartol2}. Parameter space exploration is performed using the Metropolis–Hastings algorithm~\cite{Belanger:2009ti}. We impose radiative electroweak symmetry breaking (REWSB) and require the lightest neutralino to be the LSP.

Phenomenological constraints are applied as follows. Charged sparticle masses must exceed the LEP2 limits ($\gtrsim 100$ GeV)~\cite{Patrignani:2016xqp}. The light Higgs boson mass is restricted to
\begin{equation}
122~{\rm GeV} \leq m_h \leq 128~{\rm GeV},
\end{equation}
to incorporate theoretical uncertainties in the MSSM Higgs mass calculation~\cite{Allanach:2004rh, Khachatryan:2016vau}. For colored states, we impose
\begin{align}
m_{\tilde g} &\geq 2.2~{\rm TeV}, \nonumber \\
m_{\tilde t_1} &\gtrsim 1.25~{\rm TeV}, \quad
m_{\tilde b_1} \gtrsim 1.5~{\rm TeV}, \quad
m_{\tilde q} \gtrsim 2.0~{\rm TeV},
\end{align}
consistent with current LHC limits.

Flavor and rare decay constraints are also incorporated, including $B_s\to\mu^+\mu^-$~\cite{Aaij:2012nna}, $b\to s\gamma$~\cite{Amhis:2012bh}, and $B_u\to\tau\nu_\tau$~\cite{Asner:2010qj}. Explicitly, we require:
\begin{align}
0.15 &\leq 
\frac{{\rm BR}(B_u\to\tau\nu_\tau)_{\rm MSSM}}
     {{\rm BR}(B_u\to\tau\nu_\tau)_{\rm SM}}
\leq 2.41 \quad (3\sigma), \\
2.99\times 10^{-4} &\leq {\rm BR}(b\to s\gamma) \leq 3.87\times 10^{-4} \quad (2\sigma), \\
0.8\times 10^{-9} &\leq {\rm BR}(B_s\to\mu^+\mu^-) \leq 6.2\times 10^{-9} \quad (2\sigma).
\end{align}

Finally, the relic density of the lightest neutralino is constrained by the Planck 2018 measurement~\cite{Akrami:2018vks}, within a $5\sigma$ range:
\begin{equation}
0.114 \leq \Omega_{\rm CDM} h^2 \leq 0.126.
\end{equation}
Together, these conditions delineate the phenomenologically viable GmSUGRA parameter space probed in this work.

\begin{figure}[h!]
 \centering \includegraphics[width=7.90cm]{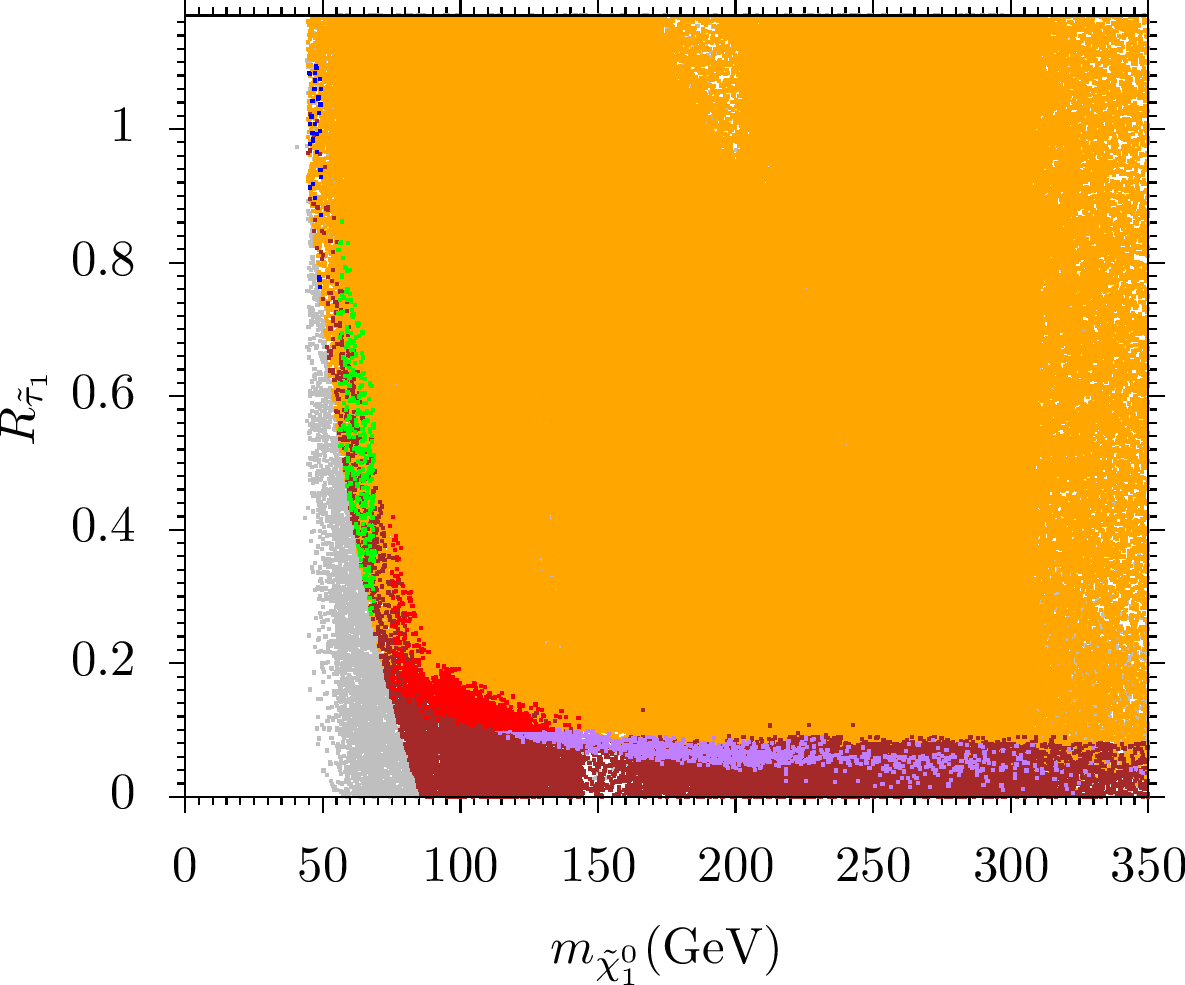}
	\caption{\small The gray points represent parameter choices that achieve radiative electroweak symmetry breaking (REWSB), yielding a neutralino as the LSP. The colored points—orange, brown, blue, green, red, and purple—represent subsets of the grey points that additionally respect LEP bounds, B-physics constraints, Higgs mass limits, and LHC sparticle search limits. Specifically, the orange and brown points corresponds to over-saturated and under-saturated dark matter relic densities, respectively. The blue, green, red, and purple regions denote $Z$-pole, Higgs-pole, bulk, and coannihilation scenarios, all consistent with the observed dark matter relic density. In this context, the mass splitting for the stau is quantified by $\mathcal{R}_{\tilde{\tau}_1} \equiv (m_{\tilde{\tau}_1}-m_{\tilde{\chi}_1^0})/m_{\tilde{\chi}_1^0}$.}
\label{F1}
\end{figure}

\begin{figure}[h!]
	\centering \includegraphics[width=8.90cm]{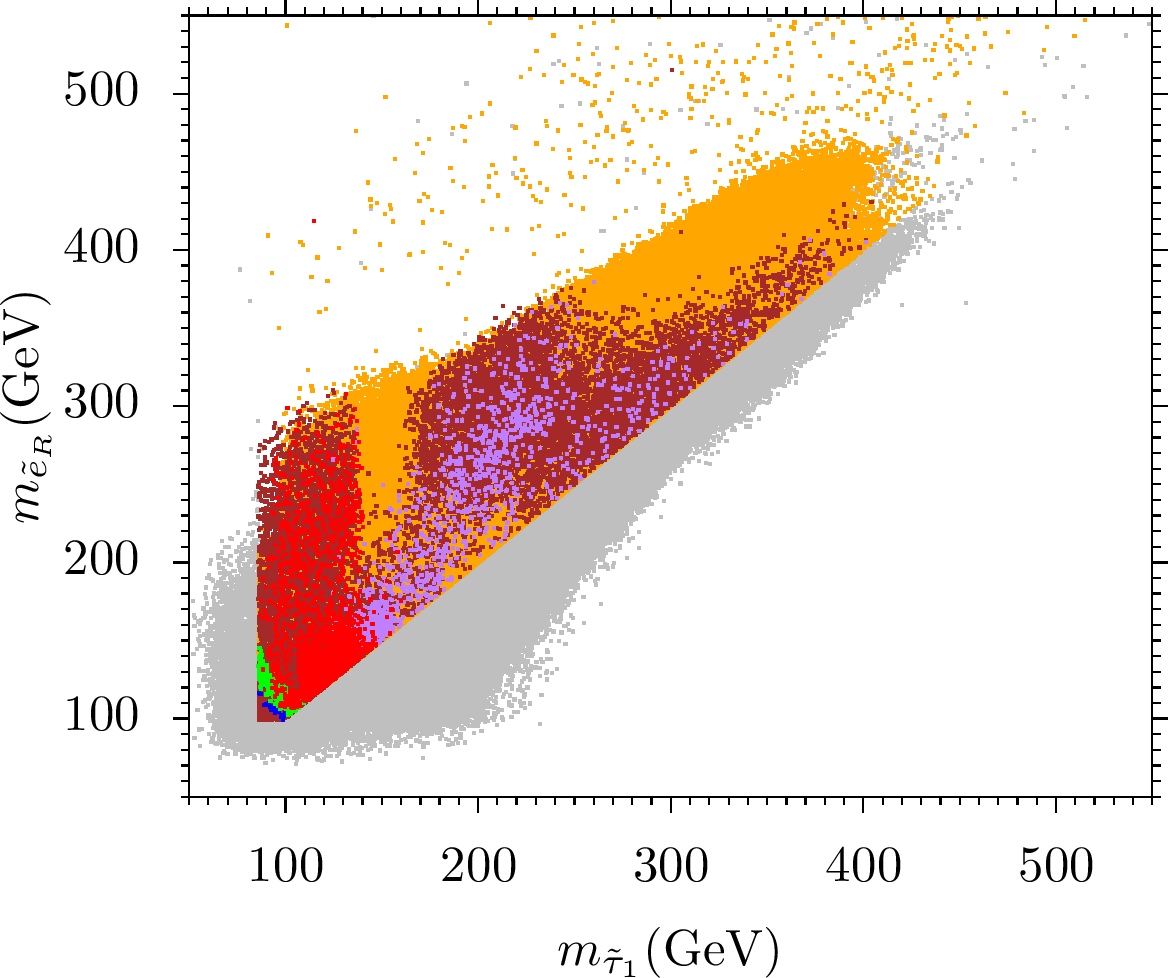}
	\caption{\small The gray points represent parameter choices that achieve radiative electroweak symmetry breaking (REWSB), yielding a neutralino as the LSP. Colored points are subsets of the grey points that additionally respect LEP bounds, B-physics constraints, Higgs mass limits, and LHC sparticle search limits. Specifically, the orange and brown points denote over-saturated and under-saturated dark matter relic densities, respectively. The blue, green, red, and purple regions correspond to $Z$-pole, Higgs-pole, bulk, and coannihilation mechanisms, each consistent with the observed dark matter relic density.}
\label{F2}
\end{figure}

\begin{figure}[h!]
        \centering \includegraphics[width=8.90cm]{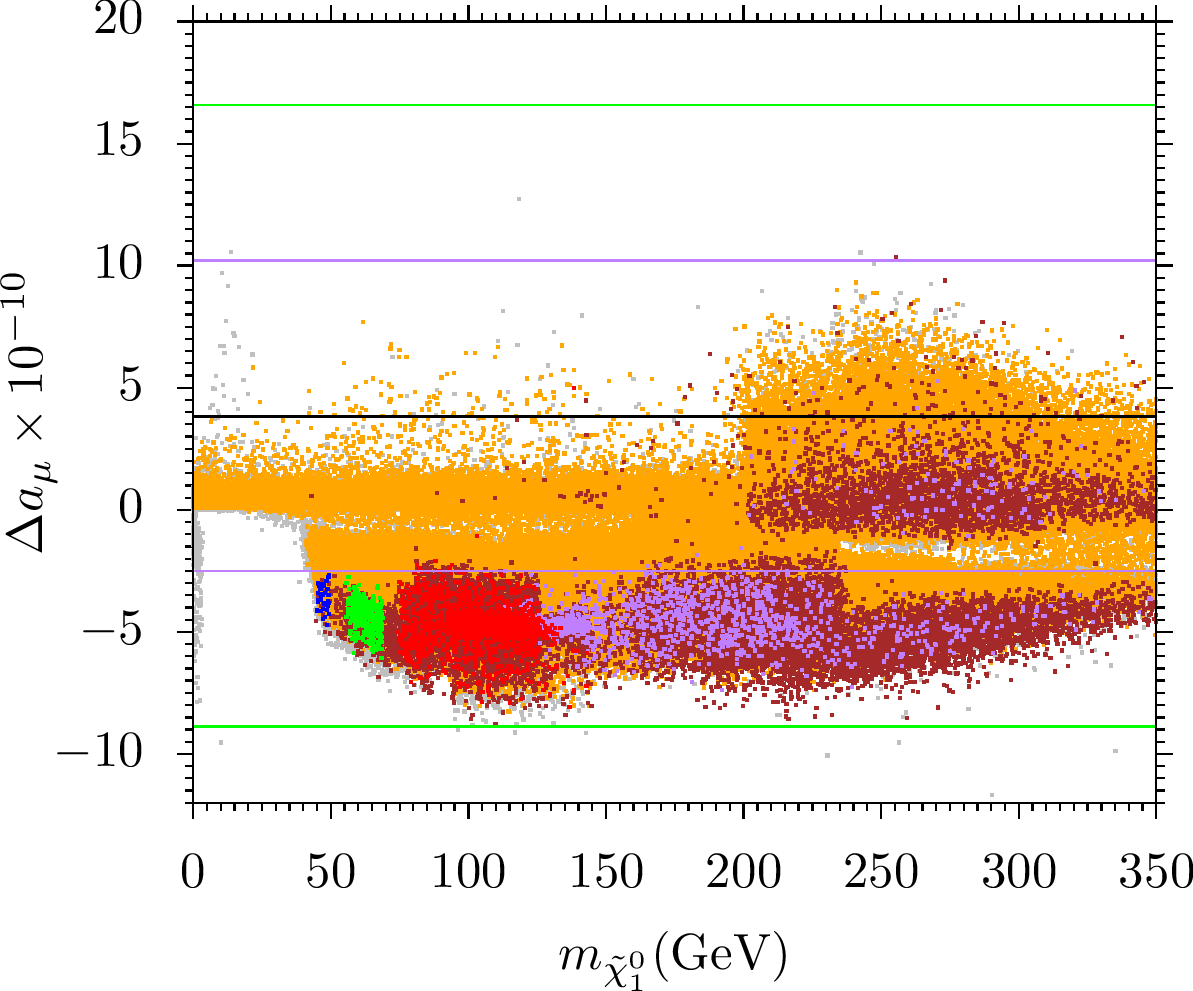}
	\caption{\small The gray points represent parameter choices that achieve radiative electroweak symmetry breaking (REWSB), yielding a neutralino as the LSP. Colored points are subsets of the grey points that additionally respect LEP bounds, B-physics constraints, Higgs mass limits, and LHC sparticle search limits. Specifically, the orange and brown points indicate over-saturated and under-saturated dark matter relic density, respectively. The blue, green, red, and purple regions correspond to $Z$-pole, Higgs-pole, bulk, and coannihilation mechanisms, all consistent with the observed dark matter relic density. The black line denotes the central value of $\Delta a_{\mu}$, while the green and purple lines indicate the $1\sigma$, and $2\sigma$ deviations, respectively.}
\label{F3}
\end{figure}

\begin{figure}[h!]
	\centering \includegraphics[width=8.7cm]{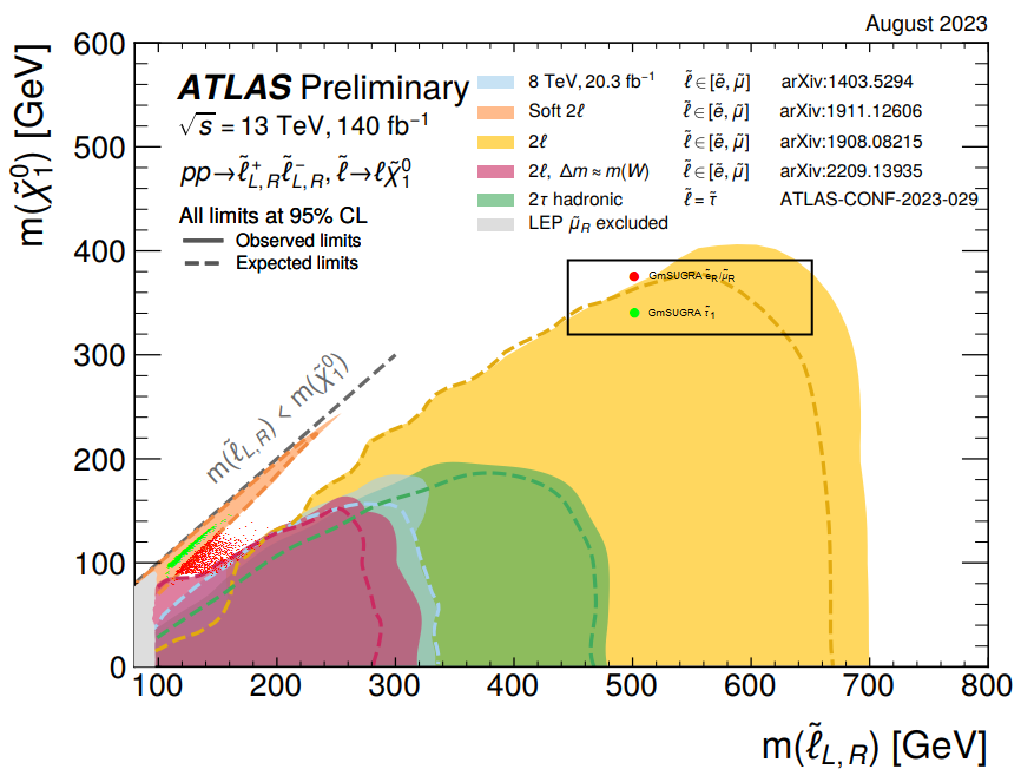}
	\caption{\small The bulk region predicted by the GmSUGRA model is superimposed on the ATLAS Collaboration’s August 2023 summary plots \cite{ATLAS:2023xco} which present updated limits on electroweak slepton production\cite{ATLAS:2014zve,ATLAS:2019lng,ATLAS:2019lff,ATLAS:2022hbt,ATLAS:2023djh}. In the GmSUGRA framework, the red points denote scenarios with light first- and second-generation right-handed sleptons, specifically a nearly degenerate pair of right-handed selectrons ($\tilde{e}_R$) and smuons ($\tilde{\mu}_R$). The green points, in contrast, correspond to configurations featuring light staus ($\tilde{\tau}_1$). All viable points lie within the annihilation-dominated regime, satisfying both the condition that the stau co-annihilation contribution is at least $\mathcal{R}{\tilde{\tau}_1}\gtrsim10\%$ and the requirement of a dark matter relic density consistent with cosmological observations.It should be emphasized that the ATLAS exclusion region shown in orange specifically targets the first two generations of sleptons namely, the selectrons 
($\tilde{e}_{L, R}$) and smuons ($\tilde{\mu}_{L, R}$), and does not apply to the stau ($\tilde{\tau}_1$), which may have different production cross-sections and decay signatures due to mixing and mass effects. The $\tilde{\tau}_1$ constraints are represented by the green-shaded region in the ATLAS plots, while the GmSUGRA points lie comfortably beyond this constraint.}
	\label{F4}
\end{figure}

\begin{figure}[h!]
        \centering \includegraphics[width=8.90cm]{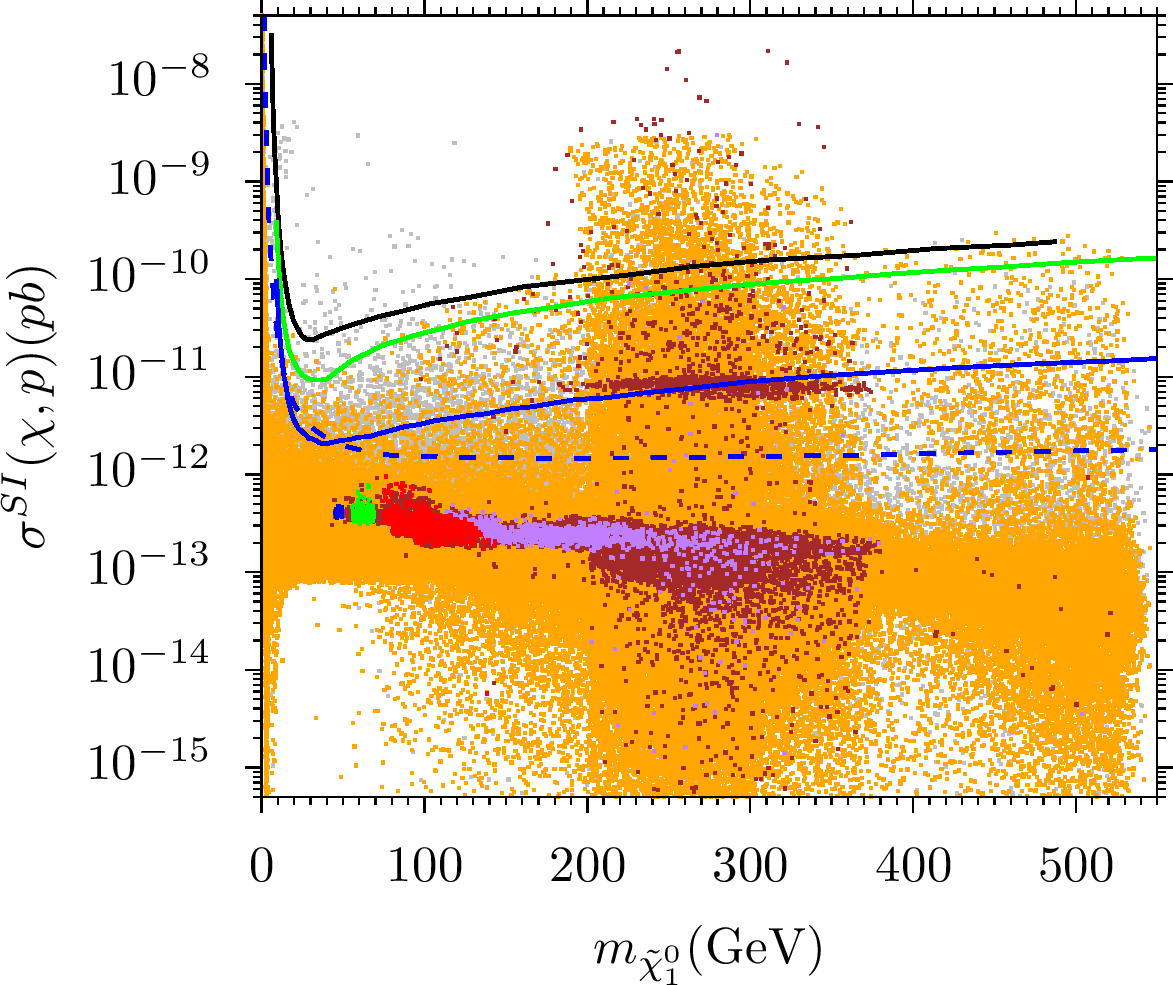}
         \centering \includegraphics[width=8.90cm]{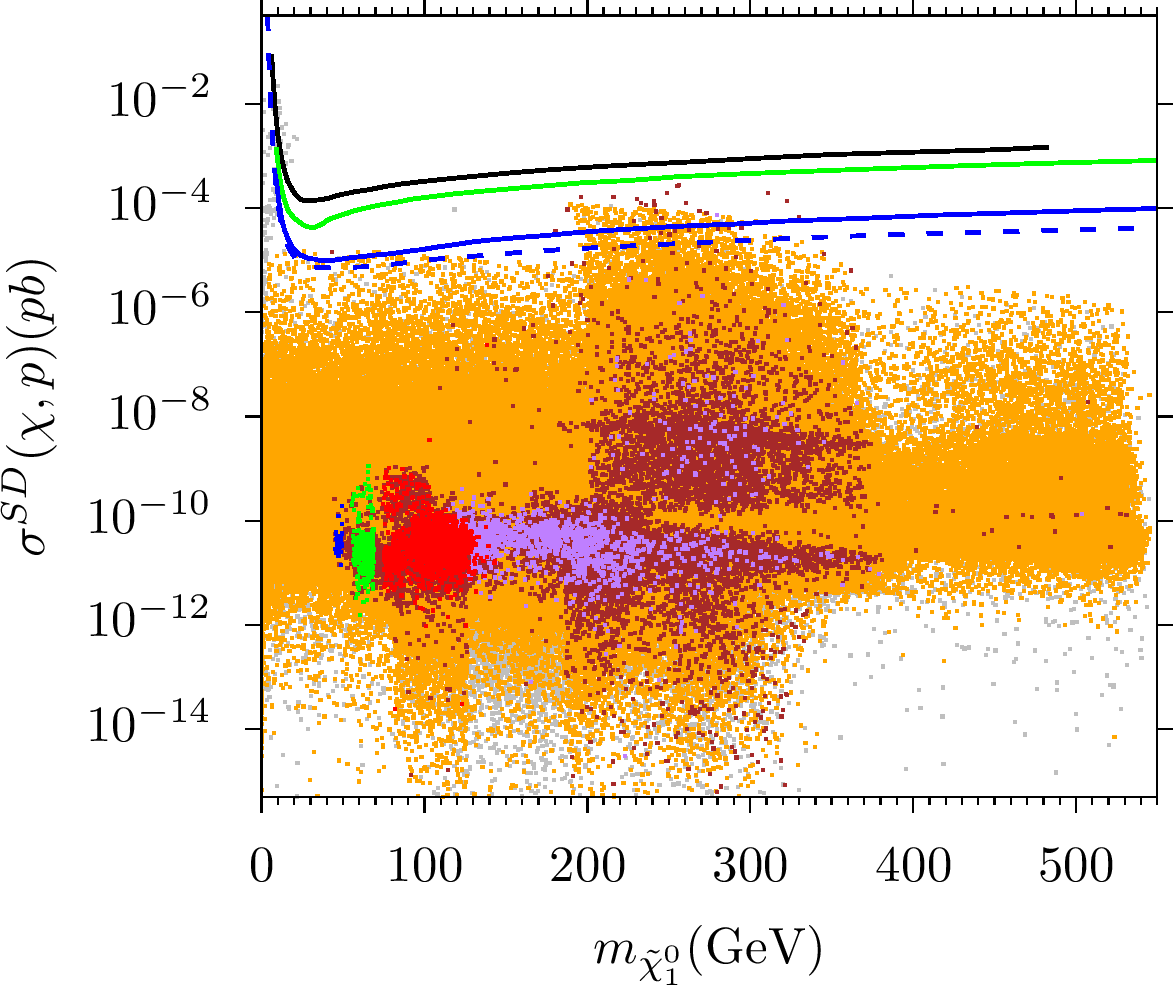}
	\caption{\small  The gray points represent parameter choices that achieve radiative electroweak symmetry breaking (REWSB), yielding a neutralino as the LSP. Colored points are subsets of the grey points that additionally respect LEP bounds, B-physics constraints, Higgs mass limits, and LHC sparticle search limits. Specifically, the orange and brown points indicate scenarios with over-saturated and under-saturated dark matter relic density, respectively. The blue, green, red, and purple regions denote parameter space consistent with saturated dark matter relic density via $Z$-pole, Higgs-pole, bulk, and coannihilation mechanisms. "The upper (lower) panel displays the spin-independent (spin-dependent) scattering cross-section between neutralinos and protons as a function of neutralino mass, overlaid with current experimental limits from XENONnT (solid black curve) \cite{XENON:2023cxc} and LZ (solid green line for 2022 data, solid blue line for 2024 sensitivity). The projected sensitivity of the 1000-day LZ run is indicated by the dotted blue line \cite{LZ:2018qzl, LZ:2022lsv,LZ:2024zvo}.}
\label{F5}
\end{figure}

\section*{Results and Discussions} 
 
\begin{table}[h!]
	\centering
	\scalebox{0.6}{
		\begin{tabular}{|l|cccc|}
			\hline
			& Point 1 & Point 2 &Point 3& Point 4    \\
			\hline
			$m_{0}^{U}$          &  170.7     & 152.3& 169.9& 190.6  \\
			$M_{1},M_{2},M_{3} $         &   277.3,-785.8,1879      & 264.6, -873.7, 1972&1159.1, -1095.5, 2041&131.1, -1188.8, 2111\\
			$m_{E^c},m_{L}$      &   137.5,682.4    &178, 933.8& 170.7, 657.4& 211.3, 557.8  \\
			$m_{H_{u}},m_{H_{d}}$           &    1035,203.5     &  1555, 218.3& 1053, 333.9&1033, 350 \\
			$m_{Q},m_{U^{c},m_{D^{c}}}$    & 160.3,191.4,0 & 156.9, 132.4, 0&170, 169.4, 0&194.2, 175.4, 0 \\
			$A_{t}=A_{b},A_{\tau}$            &    -6804,-1602     & -6927, -1319&-7228, -2186& -5884, -2050 \\
			$\tan\beta$                      & 14.4 & 18.2 & 11.9 & 12.3\\
   $\mu$                      & -3424 & -3368.6 & -3636.8 & -3288.7\\
			\hline
			$m_h$            &  125   &126&125&124   \\
			$m_H$            &  3290    & 3082&3614 & 3302 \\
			$m_{A} $         &  3269     &  3062& 3590& 3280   \\
			$m_{H^{\pm}}$    &  3291   & 3083 & 3615 & 3303    \\
			\hline
			$m_{\tilde{\chi}^0_{1,2}}$
			& 113,707 & 107,781& 61,967& 47, 1046\\
			$m_{\tilde{\chi}^0_{3,4}}$
			& 3405,3406 &3349,3350& 3617,3619& 3280,3281 \\
			$m_{\tilde{\chi}^{\pm}_{1,2}}$
			&710,3412  &784,3357&971,3626 &1051,3285 \\
			\hline
			$m_{\tilde{g}}$  & 3945    &4125& 4260&4394 \\
			\hline $m_{ \tilde{u}_{L,R}}$
			& 3450,3420  & 3608,3572& 3739,3678& 3859,3790   \\
			$m_{\tilde{t}_{1,2}}$
			& 1638, 2680  & 1724,2753&1837,2971&2421,3272 \\
			\hline $m_{ \tilde{d}_{L,R}}$
			& 3451,3451 & 3609,3571&3740,3680&3860,3789\\
			$m_{\tilde{b}_{1,2}}$
			& 2673,3244 & 2749,3296&2964,3549&3254,3685 \\
			\hline
			$m_{\tilde{\nu}_{1}}$
			& 834      & 1079&947&933  \\
			$m_{\tilde{\nu}_{3}}$
			& 822      & 1069&932&915  \\
			\hline
			$m_{ \tilde{e}_{L,R}}$
			& 855,158   & 1092, 159&969,107&955,103 \\
			$m_{\tilde{\tau}_{1,2}}$
			& 127,838    & 121, 1079&105,947&98,933\\
			\hline
			$\Delta a_{\mu}$
			& -6.6$\times 10^{-10}$ & -4.7$\times 10^{-10}$ &-3.7$\times 10^{-10}$&-3.5$\times 10^{-10}$  
			\\
			\hline
			$\sigma_{SI}(pb)$
			& 2.3$\times 10^{-13}$ & 2.5$\times 10^{-13}$&4.1$\times 10^{-13}$&4.8$\times 10^{-13}$\\
   $\sigma_{SD}(pb)$
			& 1.6$\times 10^{-11}$ & 5.7$\times 10^{-11}$ &1.4$\times 10^{-11}$&1.2$\times 10^{-10}$ 
			\\
			$\Omega_{CDM}h^2$
			& 0.123     & 0.117&0.120&0.122 \\
            	\hline
             $\mathcal{R}_{\tilde{\tau}_1}$& $12\%$   &  $12\%$&  $71\%$&  $108\%$     \\
    $\mathcal{R}_{\tilde{e}_R}$& $39\%$   &  $48\%$&  $75\%$&  $117\%$     \\
			\hline
			\hline
		\end{tabular}
  }
 \caption{The table provides the masses in the unit GeV}
		\label{table1}
\end{table}

 
A central virtue of low-energy supersymmetry lies in its natural prediction of a viable thermal relic DM candidate. In particular, the MSSM, under the assumption of $R$-parity conservation, accommodates a neutral, colorless, and stable weakly interacting massive particle (WIMP) whose relic density can be consistent with cosmological observations~\cite{Akrami:2018vks,neutralinodarkmatter,darkmatterreviews}.  

In the MSSM, the LSP arises as a linear superposition of bino, neutral wino, and higgsino states. While the wino and higgsino components interact directly with electroweak gauge bosons—leading to substantial annihilation into final states such as $f\bar{f}$, $W^+W^-$, $ZZ$, $ZH$, $hh$, and channels involving heavy Higgses ($H$, $A$, $H^\pm$)—the bino limit isolates the simplest annihilation mechanism, namely $t$-channel sfermion exchange into fermion pairs. This renders the bino-like neutralino particularly attractive for exploring the so-called \emph{bulk region} of parameter space.  

To capture this regime, we restrict ourselves to a nearly pure bino LSP ($\gtrsim 99.9\%$), thereby suppressing large annihilation cross sections from higgsino/wino admixtures and resonance channels. In this limit, the dominant process $\tilde{\chi}_1^0 \tilde{\chi}_1^0 \rightarrow f \bar{f}$ is mediated by relatively light right-handed sleptons. Unlike resonance ($A/H$-funnel) or coannihilation scenarios, the bino–slepton annihilation in the bulk region yields the correct relic density without additional fine-tuning. To remain outside Higgs- or $Z$-pole annihilation, we require $2m_{\tilde{\chi}_1^0} \ll m_{H^0}, m_{A^0}$ and $2m_{\tilde{\chi}_1^0} \gg m_h$. Coannihilation effects are minimized by enforcing the mass-splitting condition  $\mathcal{R}_{\tilde{\phi}} \equiv ({m_{\tilde{\phi}} - m_{\tilde{\chi}_1^0}})/{m_{\tilde{\chi}_1^0}} \gtrsim 10\%,$  
with $\tilde{\phi}$ denoting right-handed sleptons. In practice, we impose $\mathcal{R}_{\tilde{e}_R} \gtrsim 10\%$ and $\mathcal{R}_{\tilde{\tau}_1} \gtrsim 10\%$, and search for spectra with $m_{\tilde{\chi}_1^0} < m_{\tilde{\tau}_1} < m_{\tilde{e}_R} = m_{\tilde{\mu}_R}$. Our scan reveals that requiring $\mathcal{R}_{\tilde{\tau}_1} \gtrsim 10\%$ implies an upper bound $m_{\tilde{\chi}_1^0} \lesssim 143$ GeV. Moreover, scenarios with $\tilde{e}_R$ as the NLSP are ruled out by the ATLAS soft-lepton searches~\cite{ATLAS:2019lng}. Thus, the only viable GmSUGRA realization of the bulk region corresponds to a bino-like neutralino LSP with a moderately light $\tilde{\tau}_1$ and heavier selectron/smuon states.  

Figure~\ref{F1} illustrates the allowed points in the $\mathcal{R}_{\tilde{\tau}_1}$–$m_{\tilde{\chi}_1^0}$ plane. The red band highlights the genuine bulk solutions consistent with relic-density and collider constraints, whereas the green and blue regions correspond to Higgs- and $Z$-resonant annihilation, respectively. Thus, our analysis demonstrates that a significant portion of the viable parameter space resides in the bulk region, highlighted in red, thereby confirming the robustness of the criterion employed to delineate this scenario. In the complementary $\tilde{\tau}_1$–$\tilde{e}_R$ plane (Fig.~\ref{F2}), the maximal allowed masses within the bulk regime are $m_{\tilde{\tau}_1} \sim 158$ GeV and $m_{\tilde{e}_R} \sim 300$ GeV, underscoring the naturally light slepton–bino spectrum in this framework.  

Turning to the muon anomalous magnetic moment, $a_\mu \equiv (g-2)_\mu/2$, we recall that supersymmetric contributions are dominated by neutralino–smuon and chargino–sneutrino loops, approximately scaling as~\cite{KhalilS2017}  
\begin{equation}
\Delta a_\mu^{\text{SUSY}} \sim \frac{M_i \, \mu \, \tan\beta}{m_{\text{SUSY}}^4}, 
\end{equation}  
where $M_i$ are the electroweak gaugino masses, $\mu$ is the higgsino parameter, and $m_{\text{SUSY}}$ characterizes the slepton/electroweakino scale. A positive $\mu$ enhances $\Delta a_\mu$, rendering light sleptons in the bulk region favorable in light of the long-standing $g-2$ anomaly. However, the latest Fermilab E989 results, combined with improved lattice-QCD determinations of the hadronic vacuum polarization~\cite{Muong-2:2025xyk,Aliberti:2025beg}, show that the discrepancy between experiment and the Standard Model has effectively vanished, with the present world average $a_\mu^{\text{exp}}$ deviating from $a_\mu^{\text{SM}}$ by only $0.6\sigma$. This drastically alters the role of $(g-2)_\mu$ as a constraint on supersymmetric parameter space. Our parameter scans (Fig.~\ref{F3}) confirm that the GmSUGRA bulk solutions remain consistent with the updated $a_\mu$ determination, with deviations confined within $1\sigma$.  

Direct searches at the LHC further constrain electroweak superpartners. Figures~\ref{F4} summarize ATLAS slepton/chargino limits~\cite{ATLAS:2023xco,ATLAS:2019lff,ATLAS:2019lng,ATLAS:2022hbt}, superimposed with GmSUGRA bulk solutions. While compressed spectra push these solutions beyond current LHC reach, future $e^+e^-$ colliders such as FCC-ee~\cite{FCC:2018byv,FCC:2018evy} and CEPC~\cite{CEPCStudyGroup:2018ghi} offer strong prospects for discovery. {The sensitivity of the High-Luminosity LHC (HL-LHC) to the bulk region considered in this work remains uncertain. This is particularly relevant in light of the analysis in Ref.~\cite{CidVidal:2018eel}, which focused on a scenario with $m_{\tilde{\tau}_R} = m_{\tilde{\tau}_L}$, whereas our study considers $m_{\tilde{\tau}_R} < m_{\tilde{\tau}_L}$. A detailed investigation of this distinction and its phenomenological implications will be undertaken in future work.}  

In Fig.~\ref{F5}, we present the neutralino–proton scattering cross sections. The spin-independent ($\sigma_{SI}$) and spin-dependent ($\sigma_{SD}$) predictions, computed with \texttt{IsaTools}~\cite{Baer:2002fv}, lie safely below the current XENONnT~\cite{XENON:2023cxc} and LZ~\cite{LZ:2022lsv} bounds, but a substantial portion of the GmSUGRA bulk is expected to be probed by the projected LZ 1000-day sensitivity~\cite{LZ:2018qzl}. Thus, the bino–slepton bulk remains a robust and testable corner of the MSSM parameter space.  

Finally, in Table~\ref{table1} we summarize four representative benchmark points that encapsulate our main results. 
\textbf{Points~1 and 2} exemplify the bino–slepton bulk scenario, consistent with relic-density and collider constraints in the yet-unexplored LHC parameter space. 
In these cases, the lightest supersymmetric particle (LSP) is a bino-dominated neutralino with masses of approximately $0.113~\text{TeV}$ and $0.107~\text{TeV}$, respectively, while the next-to-lightest supersymmetric particle (NLSP) is identified as the lighter stau. 
\textbf{Point~3} illustrates the Higgs-resonant annihilation channel, featuring a bino-like neutralino LSP with mass $\sim 0.061~\text{TeV}$, whereas \textbf{Point~4} highlights the $Z$-resonance case with a bino-like LSP mass of about $0.047~\text{TeV}$. 
These benchmarks collectively capture the phenomenologically viable regimes of the GmSUGRA framework: the bino-driven bulk annihilation and the Higgs/$Z$ funnel regions.

\section*{Conclusion} 

The persistent theoretical uncertainties in the SM calculation of the muon anomalous magnetic moment, particularly in the hadronic sector, motivate an agnostic stance on the sign of the Higgsino mass parameter $\mu$. While scenarios with $\mu>0$ have traditionally been favored to reconcile the $(g{-}2)_\mu$ discrepancy, recent advances in lattice-QCD and experimental precision have diminished this tension, thereby reinstating the phenomenological relevance of the $\mu<0$ branch~\cite{Borsanyi:2020mff, Kuberski:2023qgx, CMD-3:2023rfe, Davier:2023fpl, Khan:2025azf}. Within this framework, we have demonstrated that the bino–slepton bulk region, realized for $\mu<0$, as the anomalous magnetic moment may now be consistant with the SM prediction,  provides the most natural realization of neutralino DM in the MSSM with the GmSUGRA. Our analysis identifies a well-defined parameter space characterized by light right handed sleptons and a bino-dominated LSP with $m_{\tilde{\chi}_1^0} \lesssim 143$~GeV. In this regime, the correct relic abundance is achieved via $t$-channel slepton exchange without invoking resonance or coannihilation mechanisms. The next-to-lightest supersymmetric particles is the light stau, with $m_{\tilde{\tau}_1} \lesssim 158$~GeV and $m_{\tilde{e}_R} \lesssim 300$~GeV. This “bulk” region emerges naturally from electroweak-scale supersymmetry breaking in the GmSUGRA and remains consistent with the Planck 2018 relic density determination as well as the latest constraints from XENONnT and LUX-ZEPLIN. Importantly, our results predict light sleptons in compressed spectra that evade the present LHC bounds, but may become accessible at the next-generation lepton colliders such as the FCC$_{\rm ee}$ at CERN and the CEPC in China. Additionally, the ongoing 1000-day LUX-ZEPLIN run is expected to probe a substantial portion of the identified parameter space. 

In summary, our work establishes the bino–slepton bulk region as a theoretically robust and phenomenologically viable dark matter scenario in the MSSM with the GmSUGRA. It exemplifies how supersymmetry, even under the $\mu<0$ branch, continues to offer a compelling and testable dark matter framework at the intersection of collider and astroparticle physics frontiers.

\section*{Acknowledgment} 

I.K. acknowledges support from Zhejiang Normal University through a postdoctoral fellowship under Grant No.~YS304224924. TL is supported in part by the National Key Research and Development Program of China Grant No. 2020YFC2201504, by the Projects No. 11875062, No. 11947302, No. 12047503, and No. 12275333 supported by the National Natural Science Foundation of China, by the
Key Research Program of the Chinese Academy of Sciences, Grant No. XDPB15, by the Scientific Instrument Developing Project of the Chinese Academy of Sciences, Grant No. YJKYYQ20190049, and by the International Partnership Program of Chinese Academy of Sciences for Grand Challenges, Grant No. 112311KYSB20210012.



\begin{thebibliography}{0}%
\makeatletter
\providecommand \@ifxundefined [1]{%
 \@ifx{#1\undefined}
}%
\providecommand \@ifnum [1]{%
 \ifnum #1\expandafter \@firstoftwo
 \else \expandafter \@secondoftwo
 \fi
}%
\providecommand \@ifx [1]{%
 \ifx #1\expandafter \@firstoftwo
 \else \expandafter \@secondoftwo
 \fi
}%
\providecommand \natexlab [1]{#1}%
\providecommand \enquote  [1]{``#1''}%
\providecommand \bibnamefont  [1]{#1}%
\providecommand \bibfnamefont [1]{#1}%
\providecommand \citenamefont [1]{#1}%
\providecommand \href@noop [0]{\@secondoftwo}%
\providecommand \href [0]{\begingroup \@sanitize@url \@href}%
\providecommand \@href[1]{\@@startlink{#1}\@@href}%
\providecommand \@@href[1]{\endgroup#1\@@endlink}%
\providecommand \@sanitize@url [0]{\catcode `\\12\catcode `\$12\catcode `\&12\catcode `\#12\catcode `\^12\catcode `\_12\catcode `\%12\relax}%
\providecommand \@@startlink[1]{}%
\providecommand \@@endlink[0]{}%
\providecommand \url  [0]{\begingroup\@sanitize@url \@url }%
\providecommand \@url [1]{\endgroup\@href {#1}{\urlprefix }}%
\providecommand \urlprefix  [0]{URL }%
\providecommand \Eprint [0]{\href }%
\providecommand \doibase [0]{https://doi.org/}%
\providecommand \selectlanguage [0]{\@gobble}%
\providecommand \bibinfo  [0]{\@secondoftwo}%
\providecommand \bibfield  [0]{\@secondoftwo}%
\providecommand \translation [1]{[#1]}%
\providecommand \BibitemOpen [0]{}%
\providecommand \bibitemStop [0]{}%
\providecommand \bibitemNoStop [0]{.\EOS\space}%
\providecommand \EOS [0]{\spacefactor3000\relax}%
\providecommand \BibitemShut  [1]{\csname bibitem#1\endcsname}%
\let\auto@bib@innerbib\@empty
\end{thebibliography}%


\begin{thebibliography}{}


\bibitem{gaugeunification} 
S.~Dimopoulos, S.~Raby and F.~Wilczek,
Phys. Rev. D \textbf{24}, 1681-1683 (1981)
doi:10.1103/PhysRevD.24.1681;
U.~Amaldi, W.~de Boer and H.~Furstenau,
Phys. Lett. B \textbf{260}, 447-455 (1991)
doi:10.1016/0370-2693(91)91641-8;
J.~R.~Ellis, S.~Kelley and D.~V.~Nanopoulos,
Phys. Lett. B \textbf{260}, 131-137 (1991)
doi:10.1016/0370-2693(91)90980-5;
P.~Langacker,
J. Phys. G \textbf{29}, 35-48 (2003)
doi:10.1088/0954-3899/29/1/305
[arXiv:hep-ph/0102085 [hep-ph]].



\bibitem{Georgi:1974sy}
H.~Georgi and S.~L.~Glashow,
Phys. Rev. Lett. \textbf{28}, 1494 (1972)
doi:10.1103/PhysRevLett.28.1494

\bibitem{Pati:1974yy}
J.~C.~Pati and A.~Salam,
Phys. Rev. D \textbf{10}, 275-289 (1974)
[erratum: Phys. Rev. D \textbf{11}, 703-703 (1975)]
doi:10.1103/PhysRevD.10.275

\bibitem{Mohapatra:1974hk}
R.~N.~Mohapatra and J.~C.~Pati,
Phys. Rev. D \textbf{11}, 2558 (1975)
doi:10.1103/PhysRevD.11.2558

\bibitem{Fritzsch:1974nn}
H.~Fritzsch and P.~Minkowski,
Annals Phys. \textbf{93}, 193-266 (1975)
doi:10.1016/0003-4916(75)90211-0

\bibitem{Georgi:1974my}
H.~Georgi,
AIP Conf. Proc. \textbf{23}, 575-582 (1975)
doi:10.1063/1.2947450


\bibitem{neutralinodarkmatter}
H.~Goldberg,
Phys. Rev. Lett. \textbf{50}, 1419 (1983)
[erratum: Phys. Rev. Lett. \textbf{103}, 099905 (2009)]
doi:10.1103/PhysRevLett.50.1419;
J.~R.~Ellis, J.~S.~Hagelin, D.~V.~Nanopoulos, K.~A.~Olive and M.~Srednicki,
Nucl. Phys. B \textbf{238}, 453-476 (1984)
doi:10.1016/0550-3213(84)90461-9.

\bibitem{darkmatterreviews} For reviews, see
G.~Jungman, M.~Kamionkowski and K.~Griest,
Phys. Rept. \textbf{267}, 195-373 (1996)
doi:10.1016/0370-1573(95)00058-5
[arXiv:hep-ph/9506380 [hep-ph]];
K.~A.~Olive,
[arXiv:astro-ph/0301505 [astro-ph]];
J.~L.~Feng,
eConf \textbf{C0307282}, L11 (2003)
[arXiv:hep-ph/0405215 [hep-ph]];
M.~Drees,
AIP Conf. Proc. \textbf{805}, no.1, 48-54 (2005)
doi:10.1063/1.2149675
[arXiv:hep-ph/0509105 [hep-ph]];
J.~L.~Feng,
Ann. Rev. Astron. Astrophys. \textbf{48}, 495-545 (2010)
doi:10.1146/annurev-astro-082708-101659
[arXiv:1003.0904 [astro-ph.CO]].
\bibitem{Slavich:2020zjv}
P.~Slavich, S.~Heinemeyer, E.~Bagnaschi, H.~Bahl, M.~Goodsell, H.~E.~Haber, T.~Hahn, R.~Harlander, W.~Hollik and G.~Lee, \textit{et al.}
Eur. Phys. J. C \textbf{81} (2021) no.5, 450
doi:10.1140/epjc/s10052-021-09198-2
[arXiv:2012.15629 [hep-ph]].


\bibitem{ATLAS-SUSY-Search}
ATLAS Collaboration, ATLAS-CONF-2019-040.

\bibitem{Aad:2020sgw}
G.~Aad \textit{et al.} [ATLAS],
Eur. Phys. J. C \textbf{80}, no.8, 737 (2020)
doi:10.1140/epjc/s10052-020-8102-8
[arXiv:2004.14060 [hep-ex]].

\bibitem{Aad:2019pfy}
G.~Aad \textit{et al.} [ATLAS],
JHEP \textbf{12}, 060 (2019)
doi:10.1007/JHEP12(2019)060
[arXiv:1908.03122 [hep-ex]].

\bibitem{CMS-SUSY-Search-I}
CMS Collaboration, CMS PAS SUS-19-005.


\bibitem{CMS-SUSY-Search-II}
CMS Collaboration, CMS PAS SUS-19-006.
\bibitem{Ahmed:2022ude}
W.~Ahmed, I.~Khan, T.~Li, S.~Raza and W.~Zhang,
Phys. Lett. B \textbf{832} (2022), 137216
doi:10.1016/j.physletb.2022.137216
[arXiv:2202.11011 [hep-ph]].

\bibitem{Zhang:2023jcf}
W.~Zhang, W.~Ahmed, I.~Khan, T.~Li and S.~Raza,
Phys. Rev. D \textbf{110}, no.5, 055006 (2024)
doi:10.1103/PhysRevD.110.055006
[arXiv:2304.01082 [hep-ph]].

\bibitem{Khan:2023ryc}
I.~Khan, W.~Ahmed, T.~Li and S.~Raza,
Phys. Rev. D \textbf{109} (2024) no.7, 075051
doi:10.1103/PhysRevD.109.075051
[arXiv:2312.07863 [hep-ph]].
\bibitem{Akrami:2018vks} 
N.~Aghanim \textit{et al.} [Planck],
Astron. Astrophys. \textbf{641}, A1 (2020)
doi:10.1051/0004-6361/201833880
[arXiv:1807.06205 [astro-ph.CO]].
\bibitem{Drees:1992am}
M.~Drees and M.~M.~Nojiri,
Phys. Rev. D \textbf{47} (1993), 376-408
doi:10.1103/PhysRevD.47.376
[arXiv:hep-ph/9207234 [hep-ph]].

\bibitem{King:2006tf}
S.~F.~King and J.~P.~Roberts,
JHEP \textbf{09} (2006), 036
doi:10.1088/1126-6708/2006/09/036
[arXiv:hep-ph/0603095 [hep-ph]].

\bibitem{ATLAS:2020zms}
G.~Aad \textit{et al.} [ATLAS],
Phys. Rev. Lett. \textbf{125} (2020) no.5, 051801
doi:10.1103/PhysRevLett.125.051801
[arXiv:2002.12223 [hep-ex]].


\bibitem{CMS:2020bfa}
A.~M.~Sirunyan \textit{et al.} [CMS],
JHEP \textbf{04} (2021), 123
doi:10.1007/JHEP04(2021)123
[arXiv:2012.08600 [hep-ex]].


\bibitem{ATLAS:2021moa}
G.~Aad \textit{et al.} [ATLAS],
Eur. Phys. J. C \textbf{81} (2021) no.12, 1118
doi:10.1140/epjc/s10052-021-09749-7
[arXiv:2106.01676 [hep-ex]].


\bibitem{ATLAS:2021yqv}
G.~Aad \textit{et al.} [ATLAS],
Phys. Rev. D \textbf{104} (2021) no.11, 112010
doi:10.1103/PhysRevD.104.112010
[arXiv:2108.07586 [hep-ex]].


\bibitem{CMS:2022sfi}
A.~Tumasyan \textit{et al.} [CMS],
Phys. Lett. B \textbf{842} (2023), 137460
doi:10.1016/j.physletb.2022.137460
[arXiv:2205.09597 [hep-ex]].


\bibitem{ATLAS:2022yvh}
G.~Aad \textit{et al.} [ATLAS],
JHEP \textbf{08} (2022), 104
doi:10.1007/JHEP08(2022)104
[arXiv:2202.07953 [hep-ex]].



\bibitem{XENON:2023cxc}
E.~Aprile \textit{et al.} [XENON],
Phys. Rev. Lett. \textbf{131}, no.4, 041003 (2023)
doi:10.1103/PhysRevLett.131.041003
[arXiv:2303.14729 [hep-ex]].


\bibitem{LZ:2022lsv}
J.~Aalbers \textit{et al.} [LZ],
Phys. Rev. Lett. \textbf{131}, no.4, 041002 (2023)
doi:10.1103/PhysRevLett.131.041002
[arXiv:2207.03764 [hep-ex]].

\bibitem{LZ:2024zvo}
J.~Aalbers \textit{et al.} [LZ],
[arXiv:2410.17036 [hep-ex]].

\bibitem{LZ:2018qzl}
D.~S.~Akerib \textit{et al.} [LZ],
Phys. Rev. D \textbf{101}, no.5, 052002 (2020)
doi:10.1103/PhysRevD.101.052002
[arXiv:1802.06039 [astro-ph.IM]].


\bibitem{Cheng:2012np}
T.~Cheng, J.~Li, T.~Li, D.~V.~Nanopoulos and C.~Tong,
Eur. Phys. J. C \textbf{73}, 2322 (2013)
[arXiv:1202.6088 [hep-ph]].

\bibitem{Cheng:2013hna}
T.~Cheng and T.~Li,
Phys. Rev. D \textbf{88}, 015031 (2013)
[arXiv:1305.3214 [hep-ph]].

\bibitem{Li:2014dna}
T.~Li and S.~Raza,
Phys. Rev. D \textbf{91}, no.5, 055016 (2015)
[arXiv:1409.3930 [hep-ph]].
\bibitem{Li:2010xr}
T.~Li and D.~V.~Nanopoulos,
Phys. Lett. B \textbf{692}, 121-125 (2010)
[arXiv:1002.4183 [hep-ph]].


\bibitem{Balazs:2010ha}
C.~Balazs, T.~Li, D.~V.~Nanopoulos and F.~Wang,
JHEP \textbf{09}, 003 (2010)
[arXiv:1006.5559 [hep-ph]].

\bibitem{Muong-2:2023cdq}
D.~P.~Aguillard \textit{et al.} [Muon g-2],
Phys. Rev. Lett. \textbf{131}, no.16, 161802 (2023)
doi:10.1103/PhysRevLett.131.161802
[arXiv:2308.06230 [hep-ex]].

\bibitem{Ahmed:2021htr}
W.~Ahmed, I.~Khan, J.~Li, T.~Li, S.~Raza and W.~Zhang,
Phys. Lett. B \textbf{827} (2022), 136879
doi:10.1016/j.physletb.2022.136879
[arXiv:2104.03491 [hep-ph]].

\bibitem{Muong-2:2021ojo}
B.~Abi \textit{et al.} [Muon g-2],
Phys. Rev. Lett. \textbf{126}, no.14, 141801 (2021)
doi:10.1103/PhysRevLett.126.141801
[arXiv:2104.03281 [hep-ex]].

\bibitem{Khan:2025yit}
I.~Khan, A.~Muhammad, T.~Li and S.~Raza,
[arXiv:2506.18442 [hep-ph]].

\bibitem{Khan:2025azf}
I.~Khan, W.~Ahmed, T.~Li, S.~Raza and A.~Muhammad,
[arXiv:2501.12039 [hep-ph]].

\bibitem{FCC:2018byv}
A.~Abada \textit{et al.} [FCC],
Eur. Phys. J. C \textbf{79}, no.6, 474 (2019)
doi:10.1140/epjc/s10052-019-6904-3

\bibitem{FCC:2018evy}
A.~Abada \textit{et al.} [FCC],
Eur. Phys. J. ST \textbf{228}, no.2, 261-623 (2019)
doi:10.1140/epjst/e2019-900045-4

\bibitem{CEPCStudyGroup:2018ghi}
J.~B.~Guimar\~aes da Costa \textit{et al.} [CEPC Study Group],
[arXiv:1811.10545 [hep-ex]].
\bibitem{ISAJET}
H.~Baer, F.~E.~Paige, S.~D.~Protopopescu and X.~Tata,
[arXiv:hep-ph/0001086 [hep-ph]].

\bibitem{Hisano:1992jj}
J.~Hisano, H.~Murayama and T.~Yanagida,
Nucl. Phys. B \textbf{402}, 46-84 (1993)
doi:10.1016/0550-3213(93)90636-4
[arXiv:hep-ph/9207279 [hep-ph]];
Y.~Yamada,
Z. Phys. C \textbf{60}, 83-94 (1993)
doi:10.1007/BF01650433.

\bibitem{:2009ec}
   [CDF and D0],
[arXiv:0903.2503 [hep-ex]].


\bibitem{bartol2} 
I.~Gogoladze, R.~Khalid, S.~Raza and Q.~Shafi,
JHEP \textbf{06}, 117 (2011)
doi:10.1007/JHEP06(2011)117
[arXiv:1102.0013 [hep-ph]].

\bibitem{Belanger:2009ti}
  G.~Belanger, F.~Boudjema, A.~Pukhov and R.~K.~Singh,
JHEP \textbf{11}, 026 (2009)
doi:10.1088/1126-6708/2009/11/026
[arXiv:0906.5048 [hep-ph]].

H.~Baer, S.~Kraml, S.~Sekmen and H.~Summy,
JHEP \textbf{03}, 056 (2008)
doi:10.1088/1126-6708/2008/03/056
[arXiv:0801.1831 [hep-ph]].

\bibitem{Patrignani:2016xqp}
R.~L.~Workman \textit{et al.} [Particle Data Group],
PTEP \textbf{2022}, 083C01 (2022)
doi:10.1093/ptep/ptac097

\bibitem{Khachatryan:2016vau} 
  G.~Aad {\it et al.} [ATLAS and CMS Collaborations],
  JHEP {\bf 1608}, 045 (2016)
  doi:10.1007/JHEP08(2016)045
  [arXiv:1606.02266 [hep-ex]].

\bibitem{Allanach:2004rh} 
  B.~C.~Allanach, A.~Djouadi, J.~L.~Kneur, W.~Porod and P.~Slavich,
  JHEP {\bf 0409}, 044 (2004)
doi:10.1088/1126-6708/2004/09/044

 
\bibitem{Aaij:2012nna} 
  R.~Aaij {\it et al.} [LHCb Collaboration],
  Phys.\ Rev.\ Lett.\  {\bf 110}, no. 2, 021801 (2013)
 doi:10.1103/PhysRevLett.110.021801
  [arXiv:1211.2674 [hep-ex]].
 

\bibitem{Amhis:2012bh} 
  Y.~Amhis \textit{et al.} [HFLAV],
[arXiv:1207.1158 [hep-ex]].
\bibitem{Asner:2010qj} 
  D.~Asner \textit{et al.} [HFLAV],
[arXiv:1010.1589 [hep-ex]].




\bibitem{ATLAS:2019lng}
G.~Aad \textit{et al.} [ATLAS],
Phys. Rev. D \textbf{101}, no.5, 052005 (2020)
doi:10.1103/PhysRevD.101.052005
[arXiv:1911.12606 [hep-ex]].

\bibitem{ATLAS:2023xco}
 [ATLAS],
ATL-PHYS-PUB-2023-025.

\bibitem{ATLAS:2014zve}
G.~Aad \textit{et al.} [ATLAS],
JHEP \textbf{05}, 071 (2014)
doi:10.1007/JHEP05(2014)071
[arXiv:1403.5294 [hep-ex]].

\bibitem{ATLAS:2019lff}
G.~Aad \textit{et al.} [ATLAS],
Eur. Phys. J. C \textbf{80}, no.2, 123 (2020)
doi:10.1140/epjc/s10052-019-7594-6
[{arXiv:1908.08215 [hep-ex]}].

\bibitem{ATLAS:2022hbt}
G.~Aad \textit{et al.} [ATLAS],
JHEP \textbf{06}, 031 (2023)
doi:10.1007/JHEP06(2023)031
[arXiv:2209.13935 [hep-ex]].

\bibitem{ATLAS:2023djh}
 [ATLAS],
ATLAS-CONF-2023-029.

\bibitem{KhalilS2017}
Khalil, S., $\&$ Moretti, S. (2017). Supersymmetry beyond minimality: from theory to experiment. CRC Press.

\bibitem{Muong-2:2025xyk}
D.~P.~Aguillard \textit{et al.} [Muon g-2],
[arXiv:2506.03069 [hep-ex]].

\bibitem{Aliberti:2025beg}
R.~Aliberti, T.~Aoyama, E.~Balzani, A.~Bashir, G.~Benton, J.~Bijnens, V.~Biloshytskyi, T.~Blum, D.~Boito and M.~Bruno, \textit{et al.}
[arXiv:2505.21476 [hep-ph]].


\bibitem{CMS:2023qhl}
 [CMS],
CMS-PAS-SUS-21-008.
\bibitem{CMS:2022syk}
A.~Tumasyan \textit{et al.} [CMS],
Phys. Rev. D \textbf{108}, no.1, 012011 (2023)
doi:10.1103/PhysRevD.108.012011
[arXiv:2207.02254 [hep-ex]].

\bibitem{Yuan:2022ykg}
J.~Yuan, H.~Cheng and X.~Zhuang,
[arXiv:2203.10580 [hep-ex]].


\bibitem{Baer:2002fv}
H.~Baer, C.~Balazs and A.~Belyaev,
JHEP \textbf{03}, 042 (2002)
doi:10.1088/1126-6708/2002/03/042
[arXiv:hep-ph/0202076 [hep-ph]].

\bibitem{Borsanyi:2020mff}
S.~Borsanyi, Z.~Fodor, J.~N.~Guenther, C.~Hoelbling, S.~D.~Katz, L.~Lellouch, T.~Lippert, K.~Miura, L.~Parato and K.~K.~Szabo, \textit{et al.}
Nature \textbf{593}, no.7857, 51-55 (2021)
doi:10.1038/s41586-021-03418-1
[arXiv:2002.12347 [hep-lat]].


\bibitem{Kuberski:2023qgx}
S.~Kuberski,
PoS \textbf{LATTICE2023}, 125 (2024)
doi:10.22323/1.453.0125
[arXiv:2312.13753 [hep-lat]].


\bibitem{CMD-3:2023rfe}
F.~V.~Ignatov \textit{et al.} [CMD-3],
Phys. Rev. Lett. \textbf{132}, no.23, 231903 (2024)
doi:10.1103/PhysRevLett.132.231903
[arXiv:2309.12910 [hep-ex]].


\bibitem{Davier:2023fpl}
M.~Davier, A.~Hoecker, A.~M.~Lutz, B.~Malaescu and Z.~Zhang,
Eur. Phys. J. C \textbf{84}, no.7, 721 (2024)
doi:10.1140/epjc/s10052-024-12964-7
[arXiv:2312.02053 [hep-ph]].

\bibitem{CidVidal:2018eel}
X.~Cid Vidal, M.~D'Onofrio, P.~J.~Fox, R.~Torre, K.~A.~Ulmer, A.~Aboubrahim, A.~Albert, J.~Alimena, B.~C.~Allanach and C.~Alpigiani, \textit{et al.}
CERN Yellow Rep. Monogr. \textbf{7}, 585-865 (2019)
doi:10.23731/CYRM-2019-007.585
[arXiv:1812.07831 [hep-ph]].

\end{thebibliography}
\end{document}